\pdfoutput=1
\documentclass[11pt,TimesNewRoman]{article}
\usepackage{etoolbox,ragged2e}
\usepackage[noblocks,affil-it]{authblk} 
\usepackage[symbol]{footmisc}
\usepackage{xpatch,titlesec,setspace,amstext,amsmath,graphicx,geometry,comment,color,soul}

\usepackage{hyperref}
\hypersetup{
	colorlinks=true,
	linkcolor=black,
	citecolor=blue,
}

\xpatchcmd{\author}{\relax#1\relax}{\relax\detokenize{#1}\relax}{}{}

\titleformat*{\section}{\large\bfseries}
\titleformat*{\subsection}{\small\bfseries}
\titlespacing{\section}{0pt}{12pt}{6pt}
\titlespacing{\subsection}{0pt}{0pt}{0pt}

\makeatletter
\patchcmd{\@maketitle}{center}{justify}{}{}
\patchcmd{\@maketitle}{center}{justify}{}{}
\def\maketitle{{%
  
  \AB@maketitle}}
\makeatother

\usepackage[
justification=justified,
labelsep=none,
font={footnotesize,onehalfspacing},
labelfont=bf,
]{caption}

\usepackage[
backend=biber,
style=numeric-comp,
bibstyle=nature,
sorting=none,
maxbibnames=99,
]{biblatex}
\bibliography{ref}

\begin{document}
\title{Dynamic control of anapole states with phase-change alloys}

\author[\empty]{%
  \small Jingyi Tian\textsuperscript{1,2,$\dagger$}}
\author[\empty]{%
  Hao Luo\textsuperscript{1,$\dagger$}}
\author[\empty]{%
  Yuanqing Yang\textsuperscript{3,*}}
\author[\empty]{%
  Yurui Qu\textsuperscript{1,4}}
\author[\empty]{%
  Ding Zhao\textsuperscript{1}}
\author[\empty]{%
  Min Qiu\textsuperscript{1,5,*}}
\author[\empty]{%
  Sergey I. Bozhevolnyi\textsuperscript{3}}

\affil[1]{State Key Laboratory of Modern Optical Instrumentation, College of Optical Science and Engineering, Zhejiang
University, Hangzhou 310027, China}
\affil[2]{Department of Applied Physics, Royal Institute of Technology, KTH, 10691 Stockholm, Sweden}
\affil[3]{SDU Nano Optics, University of Southern Denmark, Campusvej 55, DK-5230 Odense, Denmark}
\affil[4]{Department of Physics, Massachusetts Institute of Technology, Cambridge, MA 02139, USA}
\affil[5]{Institute of Advanced Technology, Westlake Institute for Advanced Study, Westlake University, 18 Shilongshan Road, Hangzhou 310024, China}

\setcounter{footnote}{+1}
\stepcounter{footnote}\footnotetext{These authors contributed equally to this work}
\addtocounter{footnote}{-2}
\stepcounter{footnote}\footnotetext{Corresponding emails: yy@mci.sdu.dk or minqiu@zju.edu.cn}

\date{} 

\maketitle


\section*{Abstract}
\setlength{\baselineskip}{20pt}
High-index dielectric nanoparticles supporting a distinct series of Mie resonances have enabled a 
new class of optical antennas with unprecedented functionalities. The great wealth of multipolar responses 
and their delicate interplay have not only spurred practical developments but also brought new insight 
into fundamental physics such as the recent observation of nonradiating anapole states in the optical 
regime. However, how to make such a colorful resonance palette actively tunable and even switchable among 
different elemental multipoles is still elusive. Here, for the first time, we demonstrate both theoretically and 
experimentally that a structured phase-change alloy Ge$_2$Sb$_2$Te$_5$ (GST) can support a diverse set of 
multipolar Mie resonances with active tunability and switchability. By harnessing the dramatic optical 
contrast ($\Delta n>2$) and the intermediate phases of the GST material, we realize continuous switching between 
a scattering bright state (electric dipole mode) and a dark state (anapole mode) in a broadband range ($\Delta \lambda > 600$ nm). 
Dynamic control of higher-order anapoles and resulting multimodal switching effects are also 
systematically investigated, which naturally make the structured GST serve as a multispectral optical switch with 
high extinction contrasts ($>$ 6 dB) and multi-level control capabilities. With all these findings, 
our study provides an entirely new design principle for realizing active nanophotonic devices. 
\par
\newpage

\setlength{\baselineskip}{20pt}
\noindent Ever since the seminal work of Mie \cite{mie1908beitrage}, light scattering by resonant small particles has attracted 
a vast amount of attention in many branches of physics \cite{bohren2008absorption}. The intention to control and manipulate light 
by fully exploiting the advantages from scattering resonances, particularly at the nanometer scale, 
has stimulated the emergence of modern nanophotonics \cite{novotny2012principles} 
and spawned a myriad of applications ranging from biochemistry to information technology \cite{koenderink2015nanophotonics}.
In this context, low-loss, high-index dielectric or semiconductor nanostructures featuring a diverse set of optical resonances
are currently in the spotlight of research as they can serve as versatile and CMOS-compatible building blocks 
for various photonic devices \cite{jahani2016all, kuznetsov2016optically, staude2017metamaterial, kruk2017functional, kivshar2018all}. 
Besides the practical advances, studies on dielectric nanoresonators have
also brought new insight into fundamental physics. Recent experimental investigations on the scattering response 
of Si nanoparticles have not only shown conventional radiant modes such as magnetic dipole (MD) or electric dipole (ED) resonances 
\cite{kuznetsov2012magnetic, evlyukhin2012demonstration, fu2013directional}, 
but also revealed the underlying physics of an intriguing scattering "dark" state, i.e., anapole state, 
characterized by a pronounced minimum in the scattering spectra and an asscoiated maximum in the near-field energy \cite{miroshnichenko2015nonradiating}. 
Such a suppressed scattering state stems from two antiphased electric and toroidal dipole moments, 
whose radiation patterns are identical to each other and thereby interfere destructively in the far field. 
This unique behavior of anapole states shortly unveil its tantalizing potential in many scenarios 
such as cloaking \cite{wei2016excitation,luk2017hybrid}, nanoscale lasers \cite{gongora2017anapole}, field enhancements \cite{zenin2017direct, yang2018anapole}, 
energy guiding \cite{mazzone2017near}, harmonic generations \cite{grinblat2016enhanced, grinblat2016efficient, shibanuma2017efficient} and 
metamaterials \cite{wu2018optical, yang2017multimode, papasimakis2016electromagnetic}.

However, despite the great wealth of optical resonances and rendered interesting phenomena, how to 
actively tune these responses and further switch among them remains a daunting challenge. 
This is because the induced near fields of dielectric nanoparticles, unlike their plasmonic counterparts, 
are mainly inside the structures and thereby only mildly sensitive to the change of external environments. 
For the same reason, the majority of research to date still focuses on passive structures, whose functionalities 
are set in during fabrication and cannot be altered afterward. Whereas there is a growing recognition of the need 
to realize active dielectric components, most of the published reports so far only display 
modest resonance shifts \cite{makarov2017light,sautter2015active,makarov2015tuning,shcherbakov2017ultrafast,rahmani2017reversible,bohn2018active,komar2018dynamic}. 
For instance, a pioneering work using liquid crystals as embedding media \cite{sautter2015active} generates a maximum spectral shift $\Delta\lambda\approx$ 
40 nm at resonance wavelength $\lambda \approx 1550$ nm, corresponding to a relative resonance tuning $\Delta\lambda/\lambda$ 
merely 2.6\%. A very recent study utilizing the thermo-optic effect of Si achieves a resonance shift
$\Delta\lambda\approx  30$ nm at wavelength $\lambda \approx 1500$ nm under an external temperature around 300 $^{\circ}$C \cite{rahmani2017reversible}. 
Indeed, given the multitude of resonances and follow-up functionalities offered by high-index dielectrics, 
tuning one or a few spectral peaks with limited ranges does not sufficiently employ all the benefits from such a fruitful playground. 
While attempts are also being made to obtain wideband tunability spanning over one linewidth \cite{iyer2017ultrawide,lewi2017ultrawide,holsteen2017purcell}, 
dynamic switching between or among different dielectric resonances is still an unexplored conundrum. 

In this work, we realize broadband and controllable switching between distinct scattering states 
by structuring a phase change alloy Ge$_2$Sb$_2$Te$_5$ (GST). Owing to its striking electrical and 
optical contrasts between amorphous and crystalline phases, GST has been widely used in commercial 
memory applications and was recently introduced into the nanophotonics community \cite{wuttig2017phase}. 
In contrast to the aforementioned tuning mechanisms such as using liquid crystals or temperature tuning, 
GST affords a different, non-volatile approach where the induced optical change remains stable even 
after the removal of external stimuli. So far, most of the research employs GST in the form of thin films functioning as surrounding media 
for metallic structures \cite{gholipour2013all,michel2013using,michel2014reversible,tittl2015switchable,
yin2015active,yin2017beam,qu2017dynamic,du2017control,de2018nonvolatile}. 
The characteristics of such configuration thus are still dominated by lossy plasmonic resonances. 
Although there are also exciting developments using GST itself as integrated optical constituents 
\cite{hosseini2014optoelectronic,rios2015integrated,wang2016optically,li2016reversible,feldmann2017calculating,}, 
detailed investigations on the fundamental optical properties of GST nanostructures are surprisingly lacking 
\cite{chu2016active,wuttig2017phase}. 
Here, we perform a thorough multipole analysis to examine the optical response of GST nanostructures both 
theoretically and experimentally. For the first time, we demonstrate that the high refractive index and the 
low loss of GST empower its nanostructures to support MD, ED and anapole states, in a similar manner as other 
enticing dielectrics such as Si and Ge. Meanwhile, the distinctive tunability of GST makes all these resonances 
actively controllable and switchable. By exploiting the intermediate phases of GST, 
we show progressive and continuous switching between scattering bright and dark states over an extremely 
broadband region ($\Delta\lambda/\lambda >$ 20 \%). Multimodal switching among higher-order ED and anapole states 
is also manifested, naturally making the investigated GST structures function as a multispectral optical switch with 
high extinction contrasts ($>$ 6 dB) as well as multi-level control abilities. Hence, 
by discovering the concealed portfolio of switchable resonances in GST nanostructures, our findings establish 
a new basis for designing active optical components and pave the way towards "metadevices" with 
reconfigurability and tunability on demand \cite{zheludev2012metamaterials}.

\section*{Results}
\subsection*{Mie resonances in GST structures: active tunability and switchability} 
\begin{figure}[t!]
\centering \includegraphics[width = 15cm]{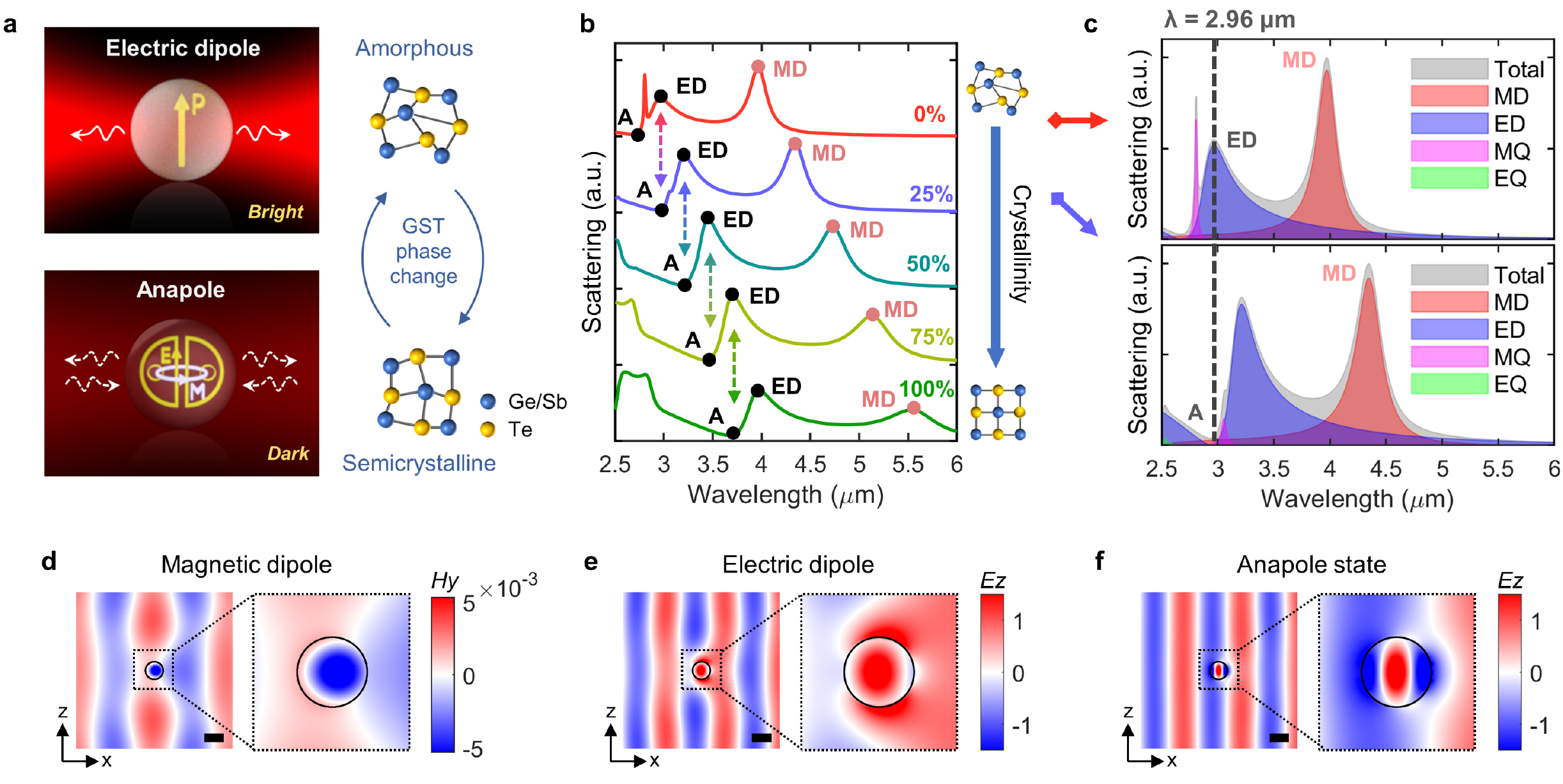}
\caption{\textbf{ $|$ Switchable scattering bright and dark states of a GST sphere.} \textnormal{(\textbf{a})
Conceptual illustration of an electric dipole resonance and an anapole excitation in a GST sphere. 
Switching between these two states can be realized by introducing an intermediate phase transition in the GST material. 
(\textbf{b}) Normalized scattering spectra of a GST sphere ($R$ = 450 nm) with different crystallinities. 
A continuous switching between ED and anapole states (denoted as A) can be clearly observed. (\textbf{c}) 
Analytical multipole analysis of two representative scattering spectra of amorphous GST (aGST) 
and 25\%-crystalline GST (25\%-cGST) spheres. (\textbf{d}, \textbf{e}) Magnetic and electric field distributions 
of the aGST sphere excited at the two scattering bright states, i.e., MD ($\lambda$ = 3.95 $\mu$m) and 
ED ($\lambda$ = 2.96 $\mu$m) resonances. (\textbf{f}) Electric field distributions of the 25\%-cGST sphere 
at the anapole excitation ($\lambda$ = 2.96 $\mu$m). The scale bars in \textbf{d}--\textbf{f} represent 1 $\mu$m.}} 
\label{Figure1}
\end{figure}

\noindent To analyze the electromagnetic response of GST nanostructures, we start our investigation by considering 
the most general case for analytical treatments: a GST sphere situated in the vacuum (see \hyperref[Methods: analytical]{Methods}). 
\hyperref[Figure1]{Fig. 1\textbf{a}} conceptually illustrates two representative scattering states, i.e., ED and anapole states, 
supported by the GST sphere. At the ED state, the induced ED moment \textbf{p} generates a considerable
dipolar radiation, giving rise to a scattering "bright" mode. By contrast, the anapole mode features an 
induced poloidal current of electric fields inside the particle, associated with a torus of circulating 
magnetic fields \cite{luk2017hybrid}. Such an intricate field distribution is characterized by a significant 
suppression of far-field scattering. Therefore this phenomenon is also commonly referred to as a "dark" mode. 
Based on the Mie theory, the spectral positions of these two modes are heavily dependent on the refractive index 
of the particle. Therefore, switching between the modes could be realized by introducing a delicate amount 
of index change to the GST sphere. To this end, thermal, electrical, and optical stimuli 
could be utilized to induce controllable phase transformations in the GST material, making it possible 
to achieve not only the transition between amorphous and crystalline states but also among intermediate 
(semicrystalline) levels \cite{wuttig2017phase}. The scattering response of a GST sphere with a fixed radius 
$R = 450$ nm and varying crystallinities $C$ is depicted in \hyperref[Figure1]{Fig. 1\textbf{b}}. With a stepped 
phase change $\Delta C = 25\%$, progressive switching between scattering maxima and minima can be readily observed. 
A detailed multipole analysis further shows that the scattering maxima and minima are ambiguously attributed 
to the ED and anapole states, respectively (\hyperref[Figure1]{Fig. 1\textbf{c}}). In particular, 
the anapole state (denoted as A) corresponds to the minimum partial scattering contributed by the spherical ED, 
as unraveled by ref \cite{miroshnichenko2015nonradiating}. Other multipolar resonances, such as MD, 
magnetic quadrupole (MQ), and electric quadrupole (EQ) modes are also distinctly manifested. The prolific 
multipolar effects of the GST sphere come from its notably high index 
($n_{\text{aGST}} > 4, n_{\text{cGST}} > 6$, see Supplementary Fig. S1), which is of central importance 
to the field of all-dielectric nanophotonics. Besides the two representative examples ($C = 0\%$ and $C = 25\%$) 
shown in \hyperref[Figure1]{Fig. 1\textbf{c}}, in Fig. S2\textbf{a} we provide detailed multipole analysis for 
other crystalline phases of GST, clarifying that the progressive switching and the multipolar effects indeed 
take place in the GST sphere with all different crystallinities.

In \hyperref[Figure1]{Fig. 1\textbf{d}--\textbf{f}} we plot the field distributions of three typical scattering 
states excited in the GST sphere. The incident planewave propagates along the $x$--axis with 
the electric fields polarized along the $z$--direction. At MD resonance (\hyperref[Figure1]{Fig. 1\textbf{d}}), 
the magnetic fields are concentrated around the center of the sphere with noticeable scattered fields 
while the associated electric fields are circulating around the center (Fig. S2\textbf{b}). At ED resonance (\hyperref[Figure1]{Fig. 1\textbf{e}}), 
the electric fields exhibit a dipolar response parallel to the incident polarization ($z$--axis). 
An appreciable scattering process also occurs, as shown in \hyperref[Figure1]{Fig. 1\textbf{e}}. 
However, in contrast to the two bright modes, the GST sphere seems to be transparent with imperceptible scattering 
at the anapole state (\hyperref[Figure1]{Fig. 1\textbf{f}}). 
The displayed field distribution with antiphased $E_z$ (\hyperref[Figure1]{Fig. 1\textbf{f}}) and 
two field zeros (Fig. S2\textbf{c}) along the $x$--direction is indeed the signature of an anapole excitation, 
as discussed in our previous work \cite{zenin2017direct,yang2018anapole}. 
Hence, the rich collection of active Mie resonances supported by GST spheres is revealed. We note that, given the plethora of 
phenomena caused by Mie resonances and associated multipolar effects \cite{jahani2016all, kuznetsov2016optically, staude2017metamaterial, kruk2017functional, kivshar2018all}, 
tunability and switchability thus can be directly implemented into many existing applications by utilizing GST resonators. 
Among all the possibilities, here we focus our attention on the switching between ED and anapole states.

\subsection*{Broadband switching between scattering bright and dark states} 
\begin{figure}[t!]
\centering \includegraphics[width = 10cm]{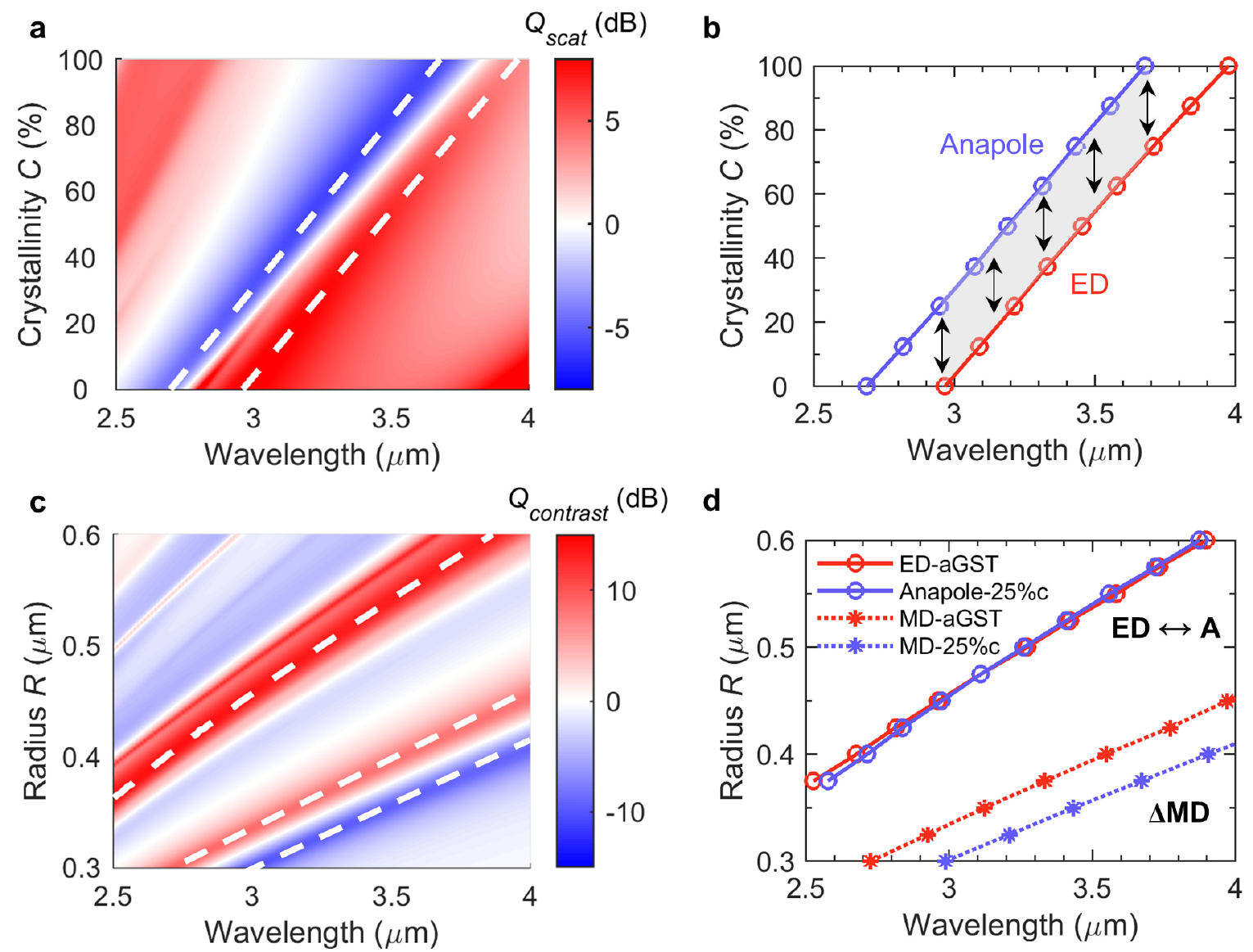}
\caption{\textbf{ $|$ Broadband tuning and switching of the multipolar scattering states in GST spheres} 
\textnormal{(\textbf{a}) Scattering efficiency $Q_{scat}$ of a single GST sphere with a fixed radius $R$ = 450 nm 
and varying crystallinities $C$. Two white dashed lines indicate the scattering maxima and minima corresponding 
to the ED and anapole states. (\textbf{b}) Analytically solved ED and anapole conditions. Switching between 
these two states can be realized within an ultrabroad bandwidth over 700 nm (the gray shaded area) by 
exploiting the intermediate levels of GST spheres. (\textbf{c}) Scattering cross-sectional contrast between aGST 
and 25\%-cGST spheres with varying radii $R$. Three white dashed lines show the most noticeable scattering contrasts. 
(\textbf{d}) Analytically solved MD, ED and anapole conditions to account for the scattering contrast between 
the aGST and 25\%c-GST spheres. The ED and anapole positions are almost perfectly overlapped in a wavelength range 
spanning around 1500 nm.}}
\label{Figure2}
\end{figure}

\noindent Next, we examine the spectral response and the bandwidth of the switching effect. Since the ED resonance and the anapole state are related 
to the maximum and the minimum of the ED contribution, they are only related to the multipole coefficient $a_1$ 
which is the function of the radius $R$, the crystallinity $C$ and the incident wavelength $\lambda$ 
(see \hyperref[Methods: analytical]{Methods} for more details). Therefore, the spectral positions of the two modes are determined by both the 
crystallinity $C$ and the geometric size $R$. In this regard, we first consider a GST sphere with an invariant radius 
($R = 450$ nm) and continously change its crystallinity $C$. The two-dimensional map of its scattering efficiency 
$Q_\text{scat}$ is plotted in \hyperref[Figure2]{Fig. 2\textbf{a}}. The pronounced scattering maxima and minima 
(marked by the white dahsed lines for eye guidance) undergo continous redshifts with the increasing crystallinity $C$. 
To precisely identify the position of ED and anapole states, we further analytically obtain the conditions for the 
two modes by solving the following sets of equations and inequalities: 
\romannumeral 1) $|a_1(\lambda,R,C)|' = 0, |a_1(\lambda,R,C)|'' < 0 $ for the ED resonance, 
and \romannumeral 2) $|a_1(\lambda,R,C)|' = 0, |a_1(\lambda,R,C)|'' > 0$ for the anapole state. With a fixed 
$R = 450$  nm, the solution set $(\lambda,C)$ is shown in \hyperref[Figure2]{Fig. 2\textbf{b}}. The shaded area 
between the ED (red) and the anapole (blue) lines represents the effective region where the GST sphere can switch 
between a ED and an anapole state. An ultra-broadband response of the switching can be observed as the area 
spans over $\Delta \lambda > 700$ nm along the $x$--axis, corresponding to a fractional bandwidth $\Delta \lambda/\lambda > 20\%$. 
Similarly, the height of the area along the $y$--axis indicates the amount of a phase change needed to 
implement the switching. Interestingly, such an amount is nearly constant over the whole spectral range, 
meaning that the presented switching functionality can be attained by simply introducing a fixed phase change 
($\Delta C \approx 25\%$) at \textit{any} given wavelengths and with \textit{any} crystalline phases of GST. 

Then we study the impacts of the geometric size on the switching effect. To this end, we consider the spheres with 
different radii $R$ and investigate the scattering contrast of these spheres at two distinct crystalline phases: 
amorphous ($C = 0\%$) and an intermediate phase ($C = 25\%$). The scattering contrast $Q_\text{constrast}$ is defined 
as the ratio between the scattering efficiencies of the spheres at the two phases. 
The 2D scattering maps of each phase are provided in Supplementary Fig. S3. 
Noticeable spectral shifts of all multipolar responses can be clearly seen when the GST spheres experience a phase 
transformation from the amorphous (Fig. S3\textbf{a}) to the semi-crystalline phase (Fig. S3\textbf{b}), or vice versa. 
Consequently, three substantial scattering contrasts over 10 dB can be found on the map of $Q_\text{constrast}$ 
(marked by the white dashed lines in \hyperref[Figure2]{Fig. 2\textbf{c}}). A multipole analysis (\hyperref[Figure2]{Fig. 2\textbf{d}}) 
further explicitly shows that these dramatic scattering contrasts are mainly attributed to two mechanisms:
\romannumeral 1) the spectral shifts of the MD resonances, giving rise to transitions between resonant and non-resonant states; 
\romannumeral 2) the switching between the ED and the anapole states, giving rise to the transition from scattering maxima to minima. 
In particular, we find that the spectral positions of the ED and the anapole states are almost perfectly overlapped with each other 
in the entire wavelength range of interest ($\Delta \lambda \approx 1500 $ nm). Therefore, given a cluster of GST spheres 
with various radii, by introducing a fixed amount of phase change (here $\Delta C = 25\%$), all the structures possessing 
different ED resonance wavelengths would exhibit the \emph{same} switching functionality from ED to corresponding anapole states. 
Such a nearly "dispersionless" switching behavior may find its applications in many interesting aspects. For instance, a major challenge 
nowadays to realizing actively tunable metasurfaces lies in the fact that metasurfaces are usually composed of meta-atoms with 
different sizes and different resonant responses. Therefore, a uniform optical change across the whole surface does not guarantee that the 
metasurface can sustain its important functions (e.g., focusing, invisibility, polarization conversion, etc) after an active tuning. 
By contrast, GST resonators with nearly dispersionless tunability and switchability may provide a promising solution to overcome 
this issue.

\titlespacing{\subsection}{0pt}{8pt}{0pt}
\subsection*{Experimental realization of switchable scattering states} 
\begin{figure}[t!]
\centering \includegraphics[width = 12cm]{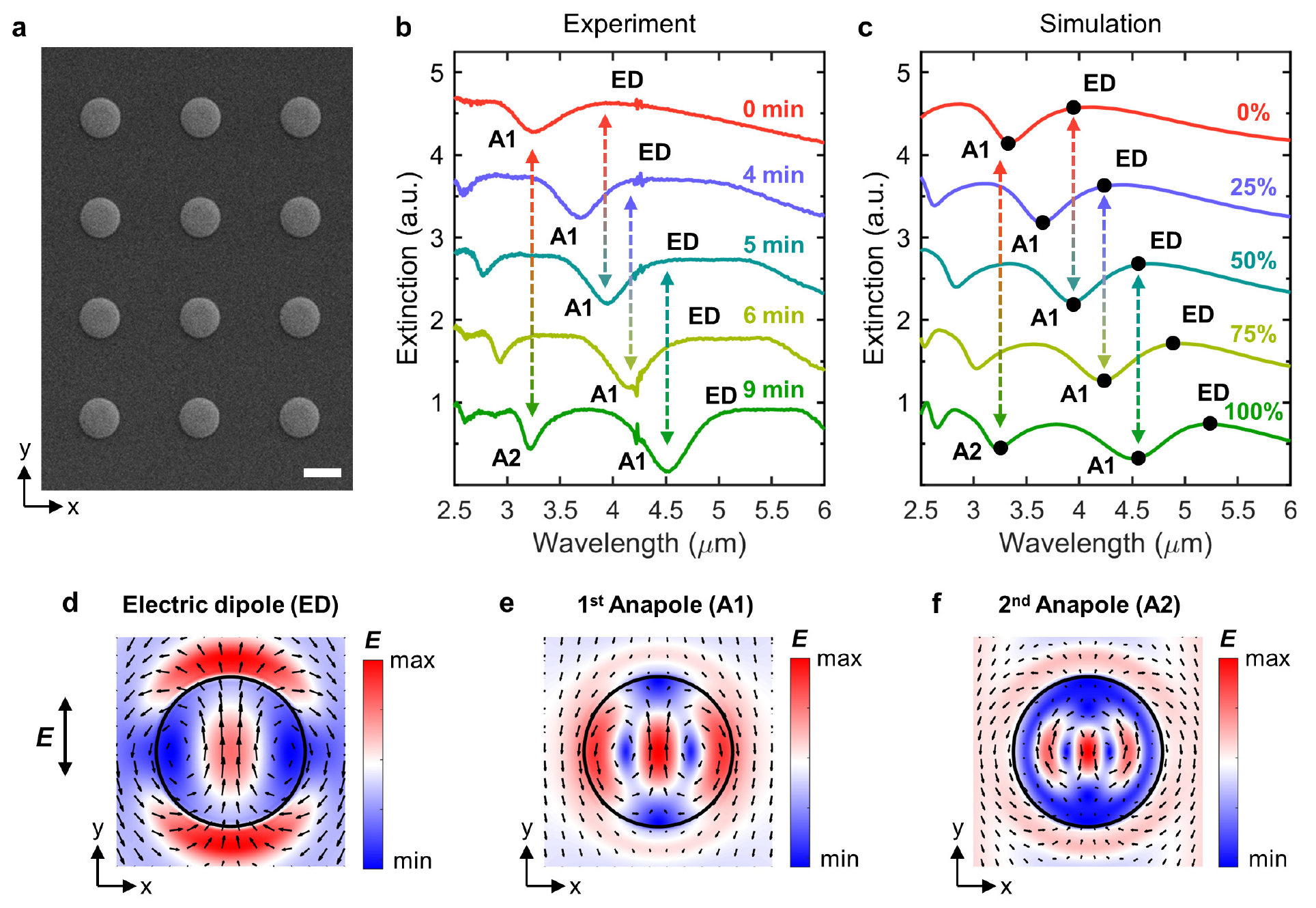}
\caption{\textbf{ $|$ Experimental realization of switchable anapole states in structured GST nanodisks} 
\textnormal{(\textbf{a}) SEM image of the fabricated GST disks. The scale bar represents 2 $\mu$m. 
(\textbf{b}, \textbf{c}) Experimental (\textbf{b}) and simulated (\textbf{c}) extinction spectra of GST disks 
with radius $R$ = 1 $\mu$m and height $H$ = 220 nm. A1 and A2 denotes the 1st-order and the 2nd-order anapole states, respectively. 
(\textbf{d--f}) Near-field distributions of the ED, A1 and A2 states in the middle planes of the 
aGST, 50\%--cGST and 100\%--cGST disks, respectively.}}
\label{Figure3}
\end{figure}

\noindent To verify the proposed concepts and predicted switching effects, we then perform experiments with GST nanostructures. 
Given the ease of fabrication and convenience for observing anapole states, truncated GST disks were fabricated 
by using E-beam lithography, magnet sputtering deposition, and standard lift-off process (see \hyperref[Methods: Sample Preparation]{Methods}). 
The geometric profile of the fabricated sample was measured by atomic force microscopy (see Fig. S4). 
The height of the disks is 220 nm and the ratio between the bottom and top radii of the GST disks is set to 2 
for all the samples based on preliminary numerical designs. Thus, in the following we can simply use the bottom radius 
$R$ to describe the geometric feature of the disks. An SEM image of fabricated GST nanodisk array with $R = 1\ \mu$m is shown in 
\hyperref[Figure3]{Fig. 3\textbf{a}}. A pitch $g$ of 3 $\mu$m was chosen to avoid coupling between adjacent nanostructures. 
The influences of the pitch size and substrate CaF$_2$ are thoroughly examined and can be referred to Fig. S5 and S6. 

To introduce different amounts of phase change and realize intermediate phases of GST, here we applied thermal stimuli by 
heating the sample on a hotplate at a fixed temperature 145$^{\circ}$C but with different amounts of time. The extinction 
spectra of the GST disks with varying heating time are presented in \hyperref[Figure3]{Fig. 3\textbf{b}}. We can clearly see 
that the GST disks indeed support notable extinction maxima and minima with evident switching between these peaks and valleys, in 
a similar manner to the GST spheres. 
Corresponding numerical results are provided in \hyperref[Figure3]{Fig. 3\textbf{c}} and an excellent agreement between the 
experimental and simulation results can be observed. Multipole decomposition of the spectral response (see \hyperref[Methods: Numerical]{Methods}) 
further clearly points out that the extinction maxima and minima are exactly correlated with the ED and anapole states, 
respectively. In particular, we find that, once a phase change of $\Delta C = 50 \%$ is introduced (e.g. from $C = 0 \%$ to $C = 50 \%$, 
$C = 25 \%$ to $C = 75 \%$, or $C = 50 \%$ to $C = 100 \%$), the GST disks would 
always undergo a switch between the ED and the first-order anapole state (A1). Such a switching functionality thus can 
be achieved in a broadband reigon over 600 nm (from 3.9 $\mu$m to 4.6 $\mu$m, corresponding to $\Delta \lambda / \lambda \approx 14\%$), 
which is remarkably consistent with our previous theroretical investigations on the GST spheres.

Besides the existence of the ED and A1 states, the large diameter-to-height ratio of the disks also enables the emergence of 
higher-order ED and anapole states, such as the second-order anapole state (A2) supported by the 100$\%$--cGST disk. 
The near-field distributions of the ED, A1 and A2 states are dipicted in \hyperref[Figure3]{Fig. 3\textbf{d}--\textbf{f}}. 
One can observe that the A2 state supports two pairs of poloidal currents which result in four field zeros along 
the $x$--axis, indicating a clear combination of the A1 state and an accompanied standing wave character. This 
phenomenon can be explained by the generation of hybrid Mie--Fabry--Perot modes \cite{yang2017multimode} or 
the superposition of several internal modes \cite{gongora2017fundamental}. Hence, the A2 state possesses 
a stronger field confinement within the disk volume compared to the A1 state, which leads to 
a higher concentration of internal energy \cite{zenin2017direct}. Therefore, such higher-order anapoles could exhibit 
their unique advantages over their fundamental counterparts, particularly in scenarios such as harmonic generation \cite{grinblat2016efficient} 
and field enhancement \cite{yang2018anapole}. Interestingly, we find that the switching between the A1 and 
the A2 states can be also realized by changing a GST disk from its amorphous phase to its crystalline phase (\hyperref[Figure3]{Fig. 3\textbf{b, c}}). 

\titlespacing{\subsection}{0pt}{8pt}{0pt}
\subsection*{Multimodal and broadband switching behavior} 
\begin{figure}[t!]
\centering \includegraphics[width = 15cm]{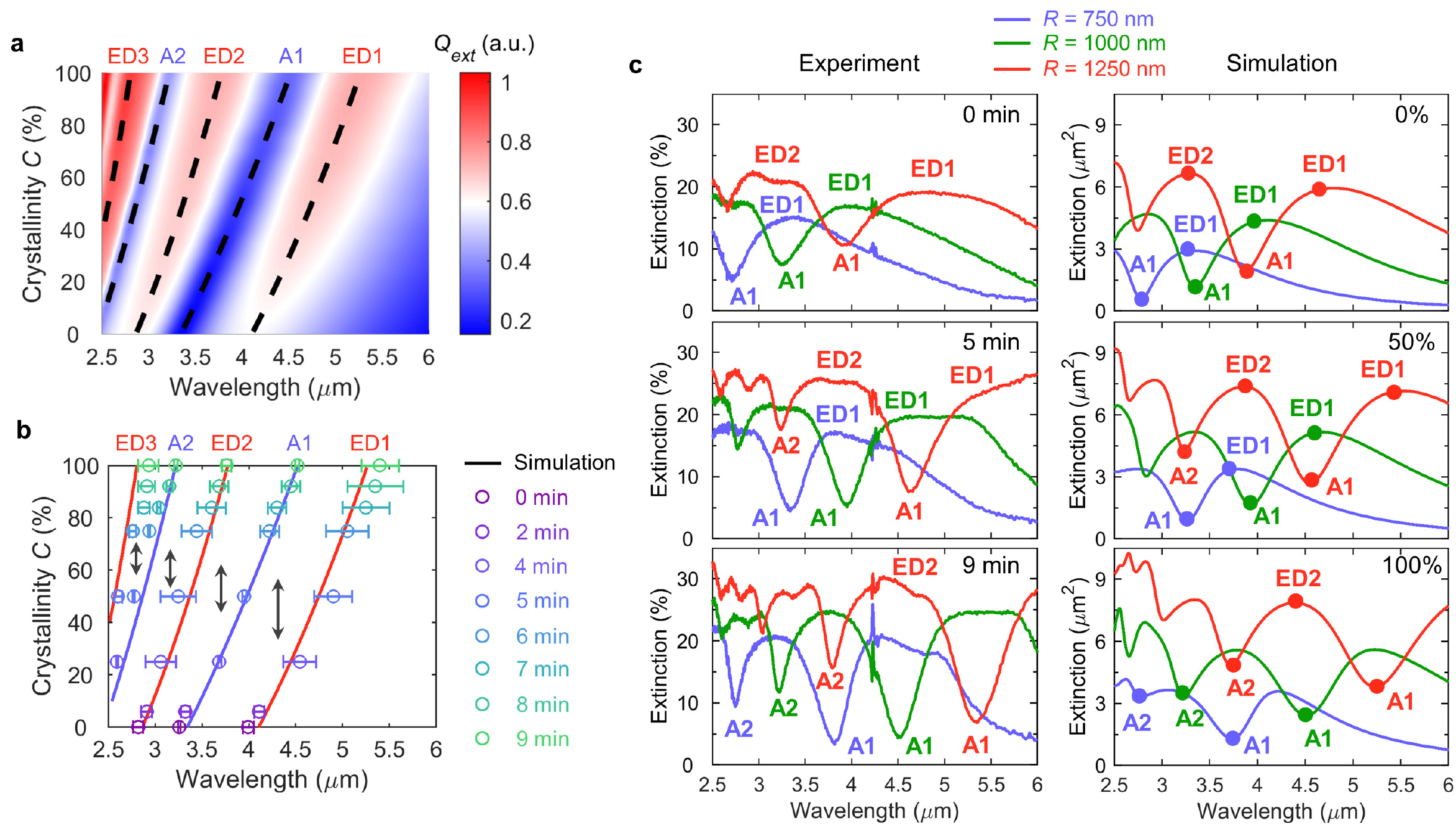}
\caption{\textbf{ $|$ Multimodal and broadband switching effects of GST disks with varying crystallinities and radius.} 
\textnormal{(\textbf{a}) Simulated 2D extinction map of GST disks with a fixed radius $R$ = 1000 nm and varying crystallinities $C$. 
The dashed lines indicate the series of bright and dark states, i.e. ED1, A1, ED2, A2, ED3. 
(\textbf{b}) Numerically solved conditions (solid lines) and experimentally measured positions (circles) 
for different bright and dark states under different crystallinities. (\textbf{c}) 
Experimental (left) and simulated (right) extinction spectra of structured GST disks with different disk radius. 
Switching functionalities such as ED to A1 or A1 to A2 are demonstrated in a broadband region over 1 $\mu$m.}}
\label{Figure4}
\end{figure}

\noindent After revealing the higher-order anapole states, we then thoroughly examine the multimodal 
response and associated switching behavior of the GST disk. In \hyperref[Figure4]{Fig. 4\textbf{a}} we 
plot the simulated 2D scattering map of the GST disk with different crystallinities $C$ varying from 0 \% 
to 100 \%. Scattering bright and dark states appear alternately across the spectra, indicating the existence 
of higher-order ED (denoted as ED2, ED3) and anapole states (see Fig. S6 for detailed multipole decomposition). 
The experimentally measured spectral positions of these states coincide well with the simulation results, as 
shown in \hyperref[Figure4]{Fig. 4\textbf{b}}. One can clearly find that, besides the demonstrated switching 
functionalities, various multimodal switching can be realized among the presented scattering states, e.g. 
the possible switching between A1 and ED2, ED2 and A2, A2 and ED3 modes, etc. It is also worth mentioning that, 
switching can not only occur between a bright and a dark state but also take place within two bright (e.g. ED2 and ED3)  
or two dark (e.g A1 to A2) states.  Compared to the GST sphere, the fabricated GST disk possesses much more fruitful 
switching phenomena due to its additional broken symmetry. Therefore one may naturally expect to unlock numerous new 
possibilities by structuring GST into different resonant shapes.

Next, we investigate GST disks with different radii $R$. Experimental and simulation extinction spectra of 
the disks at three representative crystalline phases ($C = 0\%$, $C = 50 \%$, and $C = 100 \%$) are plotted in 
\hyperref[Figure4]{Fig. 4\text{c}}. A good accordance between the experimental and simulation results can be seen 
for all the disks. In particular, when a phase change $\Delta C = 50 \%$ is introduced, all the disks exhibit the 
same switching response from the ED1 to the A1 states, despite their different ED1 resonance wavelengths $\lambda_{\text{ED1}}$ 
ranging from 3.2 $\mu$m to 4.6 $\mu$m. Similarly, when a phase change $\Delta C = 100 \%$ is introduced 
(from the amorphous phase to the crystalline phase), all the disks with different radii switch their A1 states to 
corresponding A2 states at different wavelengths $\lambda_{\text{A1}}$ spanning from 2.7 $\mu$m to 3.7 $\mu$m. Hence, 
for both aforementioned switching functionalities (ED1 to A1 and A1 to A2), nearly-dispersionless switching behaviors 
over 1 $\mu$m are demonstrated. Once again, these results substantiate our previous theoretical investigations on GST spheres. 
It is also worth noting that other multimodal responses such as the switching from A1 to ED2 and ED2 
to A2 are also sustained for all the disk sizes in a broadband region. 


\begin{figure}[h!]
\centering \includegraphics[width = 8cm]{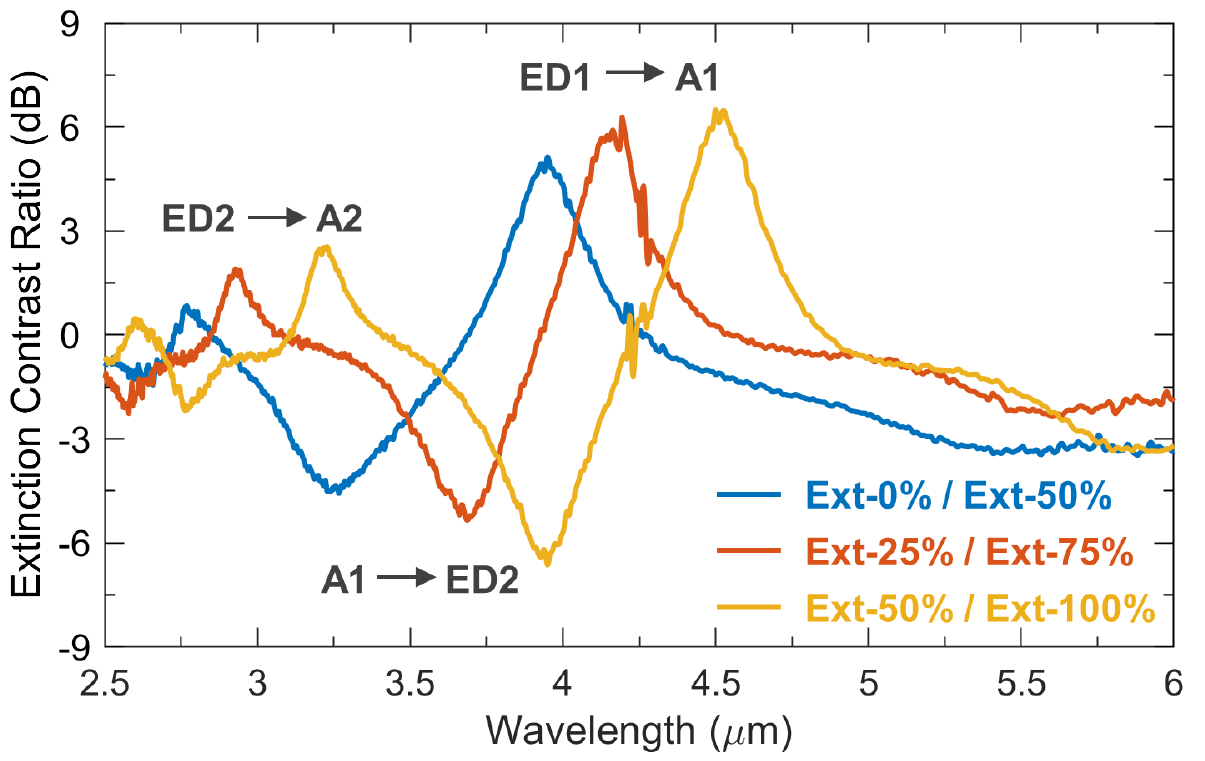}
\caption{\textbf{ $|$ Multispectral optical switch with multi-level control capabilities.} 
\textnormal{Three representative extinction contrasts of GST disks ($R = 1\ \mu$m, the same as in Fig. 3) with different crystallinities 
are plotted, which are between \romannumeral 1) c = 0\% and c = 50\% (blue), \romannumeral 2) c = 25\% and c = 75\% (red), 
and \romannumeral 3) c = 50\% and c = 100\% (yellow), respectively. }}
\label{Figure5}
\end{figure}


Finally, it is interesting to point out that the presented dynamic scattering bright and dark states
of the GST structures naturally lend themselves as an optical switch. In contrast to conventional optical switches 
which usually work at single wavelengths \cite{gholipour2013all}, here the multimodal switching 
behavior of the GST disks endow themselves with a multispectral capability. 
As we can see from Fig. 5, once a 50\% phase change is introduced (e.g., from c = 0\% to c = 50\%, the blue curve), 
the GST disks will undergo three disparate switching 
at three different wavelengths simultaneously. Both "on" and "off" states of the optical switch thus could be achieved at 
the same time in different spectral regions. A high contrast ratio of 6 dB is obtained without any 
further geometry optimization. Moreover, due to the broadband response of the switching 
functionalities, such a multispectral characteristic is sustained for different crystalline phases of the GST material (e.g. the 
red and the yellow lines in Fig. 5), which offers the possibilities for the optical switch to monitor or be controlled by 
multi-level external stimuli. Besides, as it is shown in Fig. 4\textbf{a\textnormal{,} b}, other switching functionalities 
could be also realized by exploiting other intermediate phases of the GST disks, which may find their own applications in 
many different scenarios.

\section*{Discussions}
\label{Discussion}
\noindent In summary, we investigated the fundamental optical response of individual GST nanostructures by virtue 
of a rigorous multipole analysis. We revealed that the high index and the dramatic optical contrast of 
the GST material empower its nanostructures to support a distinct series of Mie resonances with active 
tunability and switchability. Since the entire field of all-dielectric meta-optics rests on multipolar Mie resonances 
and their interference, by offering such a dynamic resonance palette, GST nanostructures thus can serve as a powerful
platform for reconfigurable nanodevices towards various applications including metasurfaces, biosensing, 
and nonlinear optics. In particular, we then demonstrated the switching between ED and anapole states, corresponding to 
scattering bright and dark modes, respectively. A broadband response of the switching effect was examined and we showed 
that the ED-to-anapole switching can be achieved in a GST nanodisk by simply introducing a phase change $\Delta C = 50 \%$ at 
any given wavelengths between 3.9 $\mu$m to 4.6 $\mu$m and with an arbitrary crystallinity of the GST disk. 
Furthermore, we demonstrated the nearly dispersionless behavior of the switching response over 1 $\mu$m, indicating that
a uniform phase change can enable structures with different sizes and different resonance wavelengths to exhibit the 
same switching functionality. In addition, the existence of higher-order ED and anapole states was presented with a thorough 
study on the multimodal switching among all these states. As a proof-of-principle application, we demonstrated a multispectral 
optical switch with multi-level control capabilities. We note that our present work provides a systematic 
yet prototypical demonstration of the active multipolar effects in GST nanostructures. Further developments 
can be implemented by utilizing electrical or optical stimuli which allows for ultrafast and reversible switching \cite{wuttig2017phase}. 
Moreover, by structuring the GST material into delicate shapes, numerous unexplored possibilities are expected to be discovered. 
Hence, we envision that our results will set up a new physical playground for active nanophotonics.


\section*{Methods}
\phantomsection
\label{Methods: analytical}
\footnotesize \setlength{\baselineskip}{16pt}
\titlespacing{\subsection}{0pt}{0pt}{0pt}
\subsection*{Analytical calculations for GST spheres.} 
\noindent To obtain the electromagnetic response of the GST spheres, we applied Mie theory \cite{bohren2008absorption}  
which offers the exact solution to the scattering problem and allows writing the scattering efficiency $Q_\text{scat}$ 
in the following simple form:
\begin{align}
Q_\text{scat} = \frac{2}{k^2R^2}\sum_{\ell=1}^{\infty}(2\ell+1)[|a_\ell|^2+|b_\ell|^2],
\end{align}
where $Q_\text{scat}$ is defined as the ratio between the scattering cross section and the geometrical cross section 
of a sphere, namely $\pi R^2$. $k = 2\pi/\lambda$ is the wave number related to the incident wavelength $\lambda$. 
The contributed multipole coefficients $a_\ell$ (electric) and $b_\ell$ (magnetic) 
can be read as:
\begin{align}
a_\ell = \frac{[D_\ell(nKR)/n + \ell/kR]\psi_\ell(kR)-\psi_{\ell-1}(kR)}{[D_\ell(nKR)/n + \ell/kR]\xi_\ell(kR)-\xi_{\ell-1}(kR)}, \\[0.5em]
b_\ell = \frac{[nD_\ell(nKR) + \ell/kR]\psi_\ell(kR)-\psi_{\ell-1}(kR)}{[nD_\ell(nKR) + \ell/kR]\xi_\ell(kR)-\xi_{\ell-1}(kR)},
\end{align}
where $n$ is the refractive index of GST. 
$D_\ell(nKR)$ is defined as $D_\ell(nKR) = \psi_{\ell}'(nkR)/\psi_\ell(nkR)$, with 
$\psi_\ell(kR)$ and $\xi_\ell(kR)$ the Riccati-Bessel functions of the first and second kind. The total 
scattering of the GST spheres was calculated by considering the multipole contributions up to the quardupole order. 

In all calculations, we adopted experimentally measured optical constants of GST (Suppmentary Fig. S1). 
For any intermediate phases with a crystallinity $C$ ($0\leq C \leq1$), 
the dielectric constant $\varepsilon_\text{GST}(\lambda, C)$ of GST can be estimated by 
using the Lorentz-Lorenz relation as follows \cite{chu2016active, aspnes1982local}:
\begin{align}
\frac{\varepsilon_\text{GST}(\lambda, C)-1}{\varepsilon_\text{GST}(\lambda, C)+2}  = C \times 
\frac{\varepsilon_\text{cGST}(\lambda)-1}{\varepsilon_\text{cGST}(\lambda)+2} + (1-C) \times 
\frac{\varepsilon_\text{aGST}(\lambda)-1}{\varepsilon_\text{aGST}(\lambda)+2},
\end{align}
where $\varepsilon_\text{aGST}$ and $\varepsilon_\text{cGST}$ are the permittivities of 
amorphous and crystalline GST, respectively. Therefore, by applying Eq. (1-4), 
we can clearly identify ED and anapole states by treating the partial scattering 
of the electric dipole: $Q_\text{scat}|_{a_1}(\lambda, R, C)$.
\phantomsection
\label{Methods: Numerical}
\footnotesize \setlength{\baselineskip}{16pt}
\titlespacing{\subsection}{0pt}{6pt}{2pt}
\subsection*{Numerical simulations and multipole decomposition} 
\noindent We performed three-dimensional FDTD simulations with a commercial software package (Lumerical). 
The optical constants of the GST disks with different crystallinities was determined by the experimental 
ellipsometric data (Fig. S1) and the equation (4). The refractive index of the substrate CaF$_2$ was set 
to 1.4. A normal-incident total-field/scattered-field planewave source was utilized to calculate the 
extinction, scattering, and absorption cross section of the GST disks. A mesh size of 10 nm was set over 
the whole volume of the GST disks. Perfectly matched layers were set as the boundaries to 
enclose the simulation area. To carry out multipole decomposition of the simulated spectra, a 
three-dimensional frequency domain field monitor was used to record the electric fields $\mathbf{E(r)}$ at 
every discretized points $\mathbf{r}$ (coordinate respective to the disk's center) inside the disks. 
By defining the polarization current $\mathbf{J(r)} = -i\omega\varepsilon_0[\varepsilon_r(\mathbf{r})-1]\mathbf{E(r)}$,  
the electric $a(\ell,m)$ and magnetic $b(\ell,m)$ spherical multipole coefficients can be calculated via 
the following formulae \cite{grahn2012electromagnetic}:
\begin{align}
a(\ell,m)  =  &\frac{(-i)^{\ell-1}k\eta}{2\pi E_0}\frac{\sqrt{(\ell-m)!}}{\sqrt{\ell(\ell+1)(\ell+m)!}} 
\int \text{exp}(-im\phi) \bigg\{ [\psi_\ell(kr)+\psi''_\ell(kr)] P_l^{m}(\text{cos}\theta)
\mathbf{\hat{r}} \cdot \mathbf{J(r)} \nonumber \\ 
&+ \frac{\psi'_l(kr)}{kr}\Big[\frac{d}{d\theta}P_l^{m}
(\text{cos}\theta)\hat{\theta}\cdot\mathbf{J(r)} - \frac{im}{\text{sin}\theta}P_l^{m}
(\text{cos}\theta)\hat{\phi}\cdot\mathbf{J(r)}\Big] \bigg\} d^3\mathbf{r},\\[0.5em]
b(\ell,m)  = &\frac{(-i)^{\ell+1}k^2\eta}{2\pi E_0}\frac{\sqrt{(\ell-m)!}}{\sqrt{\ell(\ell+1)(\ell+m)!}} 
\int \text{exp}(-im\phi)j_\ell(kr)\Big[\frac{im}{\text{sin}\theta}P_l^{m}(\text{cos}\theta)
\hat{\theta}\cdot\mathbf{J(r)} \nonumber \\ 
&+ \frac{d}{d\theta}P_l^{m}(\text{cos}\theta)\hat{\phi}\cdot\mathbf{J(r)} \Big],
\end{align}
where $E_0$ is the electric field amplitude of the incident plane wave; $\eta$ is the impedance of free space; 
$j_l(kr)$ is the spherical Bessel function of the first kind and $P_l^{m}(\text{cos}\theta)$ is the 
associated Legendre polynomials. Thus the total scattering cross section $C_{\text{scat}}$ of the GST disks 
can be written as the sum of partial contributions from these derived multipoles:
\begin{align}
C_{\text{scat}} = \frac{\pi}{k^2}\sum_{\ell=1}^{\infty}\sum_{m = -\ell}^{l}(2\ell+1)
(|a(\ell,m)|^2 + |b(\ell,m)|^2 ).
\end{align}
We note that the above equations allow for calculating spherical multipoles of arbitrarily high order. As such, 
we can unambiguously identify not only fundamental but also higher-order multipoles of GST disks. 

\phantomsection
\label{Methods: Sample Preparation}
\footnotesize \setlength{\baselineskip}{16pt}
\subsection*{Sample preparation} 
\noindent A 280-nm-thick PMMA (950K AR-P 672.11) was spun onto CaF$_2$ substrate 
as an electron beam resist and baked on a hotplate for 3 minutes at 150 $^{\circ}$C. Then a 50-nm-thick conductive 
protective coating (AR-PC 5090.02) was spun onto the PMMA film and baked for 2 minutes at 90 $^{\circ}$C. 
This coating is used for the dissipation of e-beam charges on insulating substrates. The PMMA was exposed to define 
a nanohole array by E-beam lithography. All e-beam patterning was performed by SEM, which is equipped with a 
Raith Elphy Quantum lithography system. The conductive layer was dissolved in DI water for 1 minute and then the 
PMMA was developed in the developer (AR 600-56) for 3 minutes followed by rinsing in IPA. After the development, 
a 220-nm-thick GST film was then deposited onto the sample by magnetic sputtering with 50W DC sputtering power while 
the substrate temperature was kept at room temperature. The deposited GST thin film was in its amorphous phase. 
The GST nanodisk array was realized after lift-off by ultra-sonic processing in acetone for 1 minute. 

\subsection*{Sample characterization} 
\noindent Transmission spectra were measured under normal incidence by 
using an infrared microscope (Hyperion1000) coupled to a Fourier transform infrared spectrometer (FTIR, Vert	ex70). 
The detector used in the measurement is an MCT detector integrated in the microscope. Air was used as the 
reference for determing the transmittance (T). Experimental extinction spectra were then derived as 1 - T. 
Phase tranformation of the GST material was induced by baking the samples on a 
hot plate which maintained a temperature of 145 $^{\circ}$C. To ensure a systematic optical characterization of each sample, 
the phase-changing process and tranmission measurement were implemeneted progressively. After being heated for 
1 minute, the sample was cooled down naturally and then transmission spectra were measured. Another cycle of 
the annealing process and the extinction measurement would be carried out thereafter until the GST disks were fully crystallized. 
The scanning electron microscope images were taken by Zeiss Ultra55 and the atomic force microscope(AFM) images were taken 
by VEECO Multimode.

\newpage
\AtNextBibliography{\small}
\printbibliography

\section*{Acknowledgements}
Y. Y. and S. I. B. acknowledge funding support from the European Research Council (the PLAQNAP project, Grant 341054) and the 
University of Southern Denmark (SDU2020 funding). J. T., H. L., Y. Q, and M. Q. acknowledge funding support from the 
National Key Research and Development Program of China (Grant No. 2017YFA0205700) and the National Natural Science Foundation of China 
(Grant No. 61425023). J. T., and Y. Q. were also supported by Chinese Scholarship Council (CSC No. 201600160020, and No. 201706320254). 
Y. Y. thanks Dr. Kaikai Du for helpful discussions.

\section*{Author contributions}
Y. Y. and D. Z. initiated the raw idea of phase-change resonators. 
Y. Y conceived the presented concept and designed the experiment.
J. T., Y. Y., and Y. Q. developed the theory and performed the computations. 
H. L. fabricated and characterized the samples. Y. Y. drafted the manuscript 
with J. T., and H. L. Y. Y., M. Q., and S. I. B. co-supervised the project and coordinated all the work. 
All the authors discussed the results and contributed to the final manuscript.

\section*{Additional information}
\textbf{\small Competing financial interests:} The authors declare no competing financial interests.

\end{document}


\linespread{1.6}

\title{\sf\Large\center{Supplementary Information: \\Dynamic control of anapole states with phase-change alloys}} 

\author[\empty]{%
  \normalsize\sf Jingyi Tian\textsuperscript{1,2,$\dagger$}}
\author[\empty]{%
  \sf Hao Luo\textsuperscript{1,$\dagger$}}
\author[\empty]{%
  \sf Yuanqing Yang\textsuperscript{3,*}}
\author[\empty]{%
  \sf Yurui Qu\textsuperscript{1,4}}
\author[\empty]{%
  \sf Ding Zhao\textsuperscript{1}}
\author[\empty]{%
  \sf Min Qiu\textsuperscript{1,5,*}}
\author[\empty]{%
  \sf Sergey I. Bozhevolnyi\textsuperscript{3}}

\affil[1]{\rm\it{State Key Laboratory of Modern Optical Instrumentation, College of Optical Science and Engineering, Zhejiang
University, Hangzhou 310027, China}}
\affil[2]{Department of Applied Physics, Royal Institute of Technology, KTH, 10691 Stockholm, Sweden}
\affil[3]{SDU Nano Optics, University of Southern Denmark, Campusvej 55, DK-5230 Odense, Denmark}
\affil[4]{Department of Physics, Massachusetts Institute of Technology, Cambridge, MA 02139, USA}
\affil[5]{Institute of Advanced Technology, Westlake Institute for Advanced Study, Westlake University, 18 Shilongshan Road, Hangzhou 310024, China}

\setcounter{footnote}{+1}
\stepcounter{footnote}\footnotetext{\small{These authors contributed equally to this work. \vspace{0.5mm}}}
\addtocounter{footnote}{-2}
\stepcounter{footnote}\footnotetext{\small{To whom correspondence should be addressed. \vspace{55mm} \\}}

\date{} 
\maketitle

\vspace{-17.5mm}
\center\sf{E-mail: yy@mci.sdu.dk; minqiu@zju.edu.cn}
\newpage

\flushleft \rm 
\titlespacing{\section}{0pt}{12pt}{0pt}
\section*{Optical constants of the GST material}
\noindent 
\justifying
The refractive indices of the amorphous (aGST) and the crystalline GST (cGST) thin films were experimentally determined by 
a spectrophotometric approach \cite{du2017control}. For intermediate phases of the GST material, 
their optical constants were estimated by using the Lorentz--Lorenz relation \cite{aspnes1982local} as described in the 
Methods section.
\begin{figure}[h!]
\centering \includegraphics[width = 8cm]{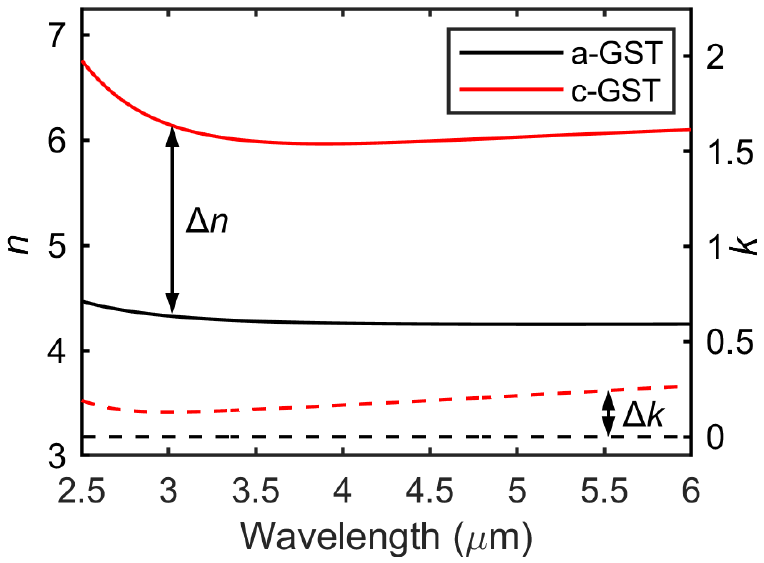}
\caption{\textbf{.} Refractive indices of the amorphous (aGST) and the crystalline GST (cGST), 
produced by magnetron sputtering deposition.}
\label{FigureS1}
\end{figure}

\newpage
\section*{Multipole expansion for GST spheres with progressive crystallinities}
\noindent 
\justifying
To examine the physical origins of the dynamic scattering bright and dark states in the GST spheres, in Fig. S2\textbf{a} we plot 
the multipole expansion of the scattering response of the GST sphere with varied crystallinities. It is observed that 
the progressive switching between the scattering maxima and minima is indeed attributed to the excitation of the ED and the anapole 
states. Fig. S2\textbf{b} provides the total electric field distribution of three representative scattering states (MD, ED, and anapole) as 
supplements to Fig. 1\textbf{d}--\textbf{f} in the main text. 
\begin{figure}[h!]
\centering \includegraphics[width = 10cm]{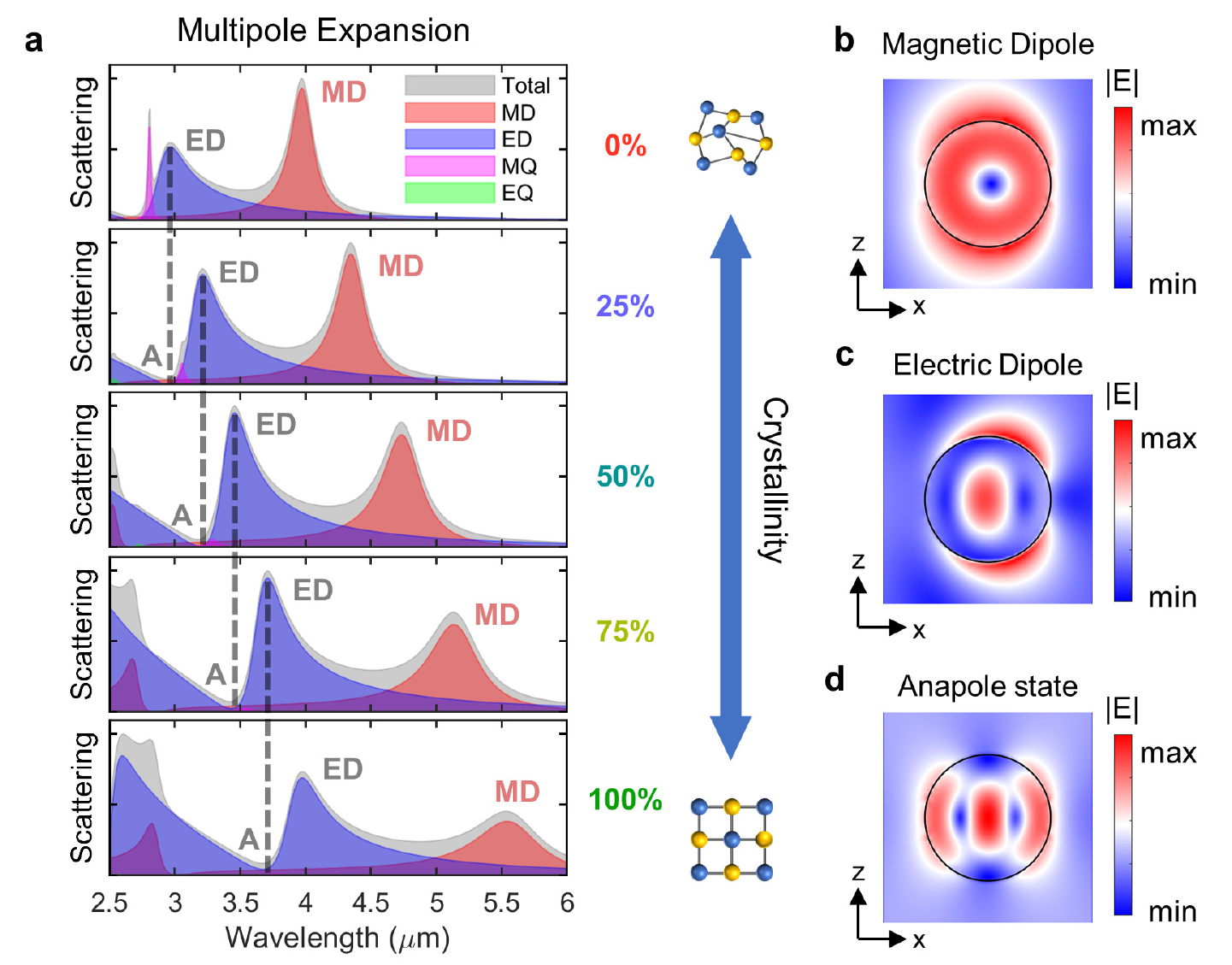}
\caption{\textbf{.} (\textbf{a}) Progressive multipole decomposition of the scattering spectra of the GST sphere in Fig. 1. 
(\textbf{b}) Total electric field distributions of the three representative scattering states, i.e. MD, ED, and the anapole state.}
\label{FigureS2}
\end{figure}

\newpage
\section*{Scattering efficiencies $Q_{scat}$ of the aGST and the 25\%-cGST spheres}
\noindent 
\justifying
To verify the "nearly--dispersionless" behavior of the switching effect, in Fig. 2c we plot the scattering contrast of the GST spheres with 
two different crystallinities, i.e. $C = 0\%$ and $C = 25\%$ and then analytically investigate the conditions for a rigorous switching (Fig. 2d).
Here, as supplements, the scattering efficiencies $Q_{scat}$ of the aGST and the 25\%-cGST spheres are provided in Fig. S3, respectively. 
\begin{figure}[h!]
\centering \includegraphics[width = 15cm]{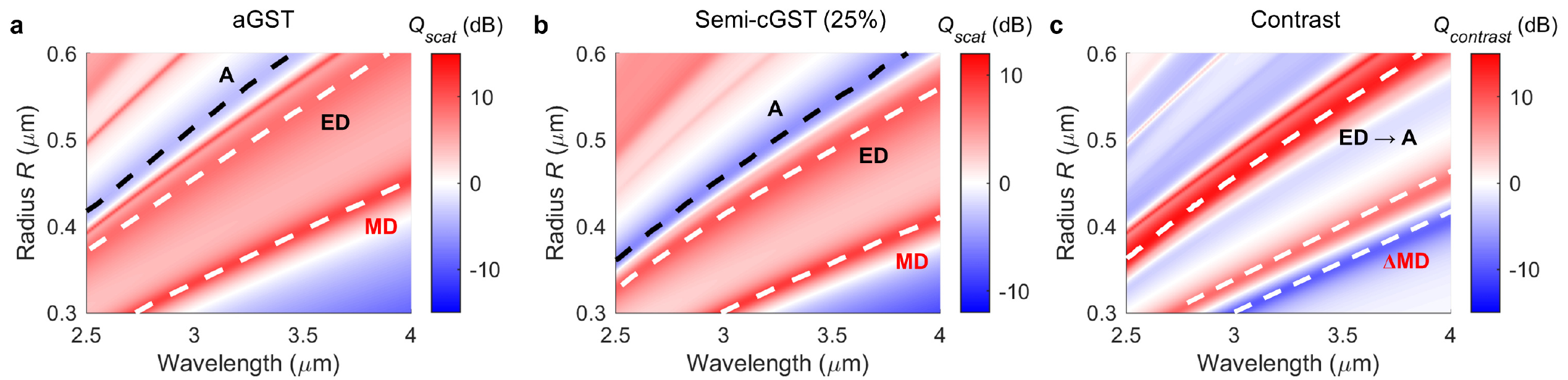}
\caption{\textbf{.} Scattering efficiencies $Q_{scat}$ of amorphous (\textbf{a}) GST spheres and 25\%-cGST (\textbf{b}) with varying radii R. 
(\textbf{c}) The scattering cross sectional contrast of GST spheres at the two phases, which is defined as $Q_{contrast} = Q_{scat-\text{aGST}}/Q_{scat-\text{25\%-cGST}}$.  
We mention that the Fig. S3 is the same as Fig. 2c and it is provided here just for ease of reference.}
\label{FigureS3}
\end{figure}

\newpage
\section*{AFM measurement of the disks' geometric profile}
\noindent 
\justifying
\vspace{-8mm}
\begin{figure}[h!]
\centering \includegraphics[width = 12cm]{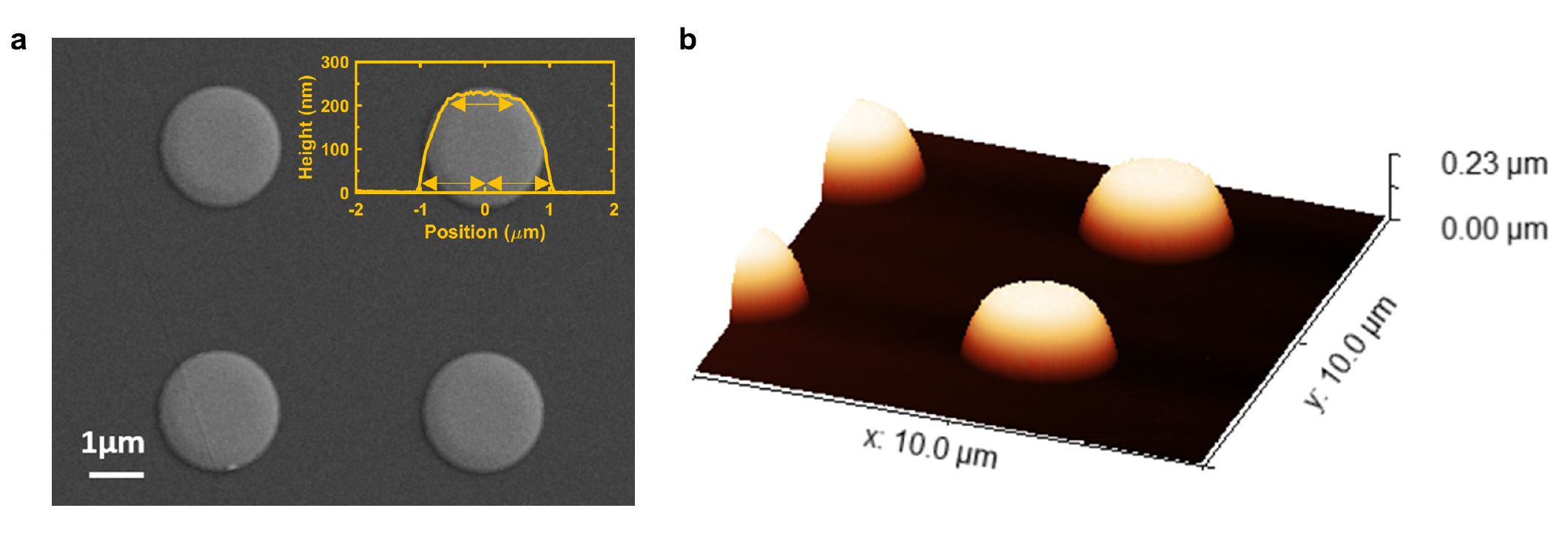}
\caption{\textbf{.} Geometric profiles of the fabricated GST nanodisks. (\textbf{a}) A zoom-in SEM image as a supplement to Figure 3a in the main text. 
The inset shows the AFM data of the cross section of the GST disk. The ratio between the bottom and the top radius was measured as 2, 
the same as devised in preliminary numerical designs. 
(\textbf{b}) 3D AFM data of a fabricated array of GST disks.}
\label{FigureS4}
\end{figure}

\newpage
\section*{Influences of the pitch size, the absorption, and the substrate on the scattering states}
\noindent 
\justifying
In this supplementary section, we discuss the impacts of the pitch size (inter-particle distance), the absorption of the GST material, 
and the existence of the substrate on the investigated scattering states and multipolar responses.

\begin{figure}[h!]
\centering \includegraphics[width = 12cm]{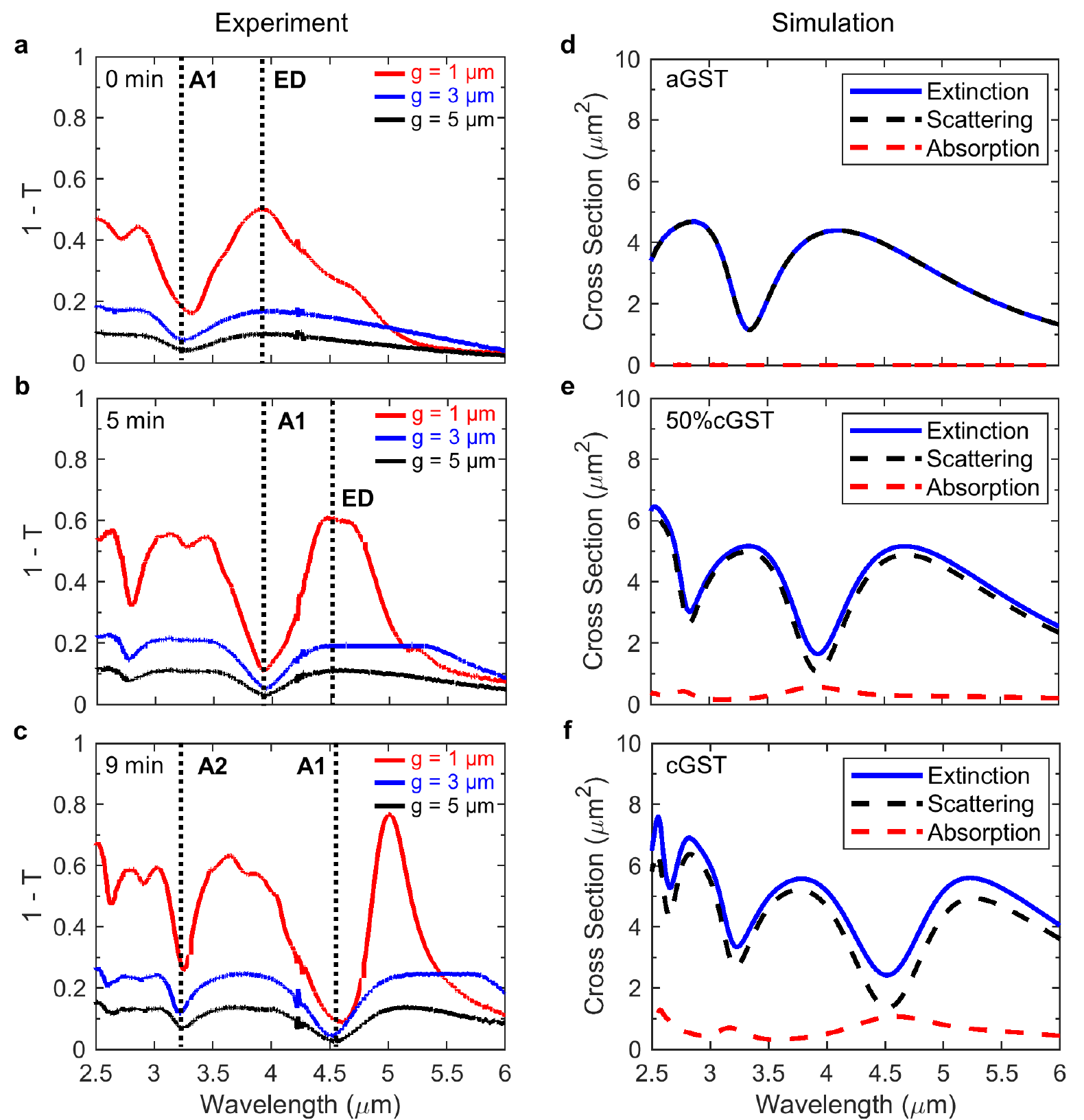}
\caption{\textbf{.} (\textbf{a}--\textbf{c}) Extinction spectra of the GST disks with a bottom radius $R$ = 1 $\mu$m and different pitch sizes $g$.  
(\textbf{d}--\textbf{f}) Extinction, scattering and absorption cross section spectra of individual GST disks ($R$ = 1 $\mu$m) at three different crystalline phases. The influence of the absorption is mild in all cases and the extinction features (pronounced peaks and dips) correlates well with those on the scattering spectra. }
\label{FigureS5}
\end{figure}
As shown in Fig. S5\textbf{a}--\textbf{c}, for different pitch sizes $g$, the optical coupling 
between adjacent GST disks does not strongly influence the spectral positions of the investigated ED and anapole states. In particular, there is generally 
no difference between the case of $g = 3\ \mu$m and $g = 5\ \mu$m. In the main text, to highlight the resonances of the disks and to mitigate the influence of 
the CO$_2$ absorption (around 4.3 $\mu$m), we used the spectra with $g = 3\ \mu m$. In Fig. S5\textbf{d}--\textbf{f}, we provide the scattering , the absorption, and 
the extinction spectra of individual GST disks with different crystallinities. Indeed, as the crystallinity increases, the absorption of the GST material increases. 
However, given the relatively low loss of the GST material (Fig. S2) in the wavelength range of interest, the influence of the absorption in mild and the 
extinction features (pronounced peaks and dips) correlate well with those on the scattering spectra.

\begin{figure}[h!]
\centering \includegraphics[width = 12cm]{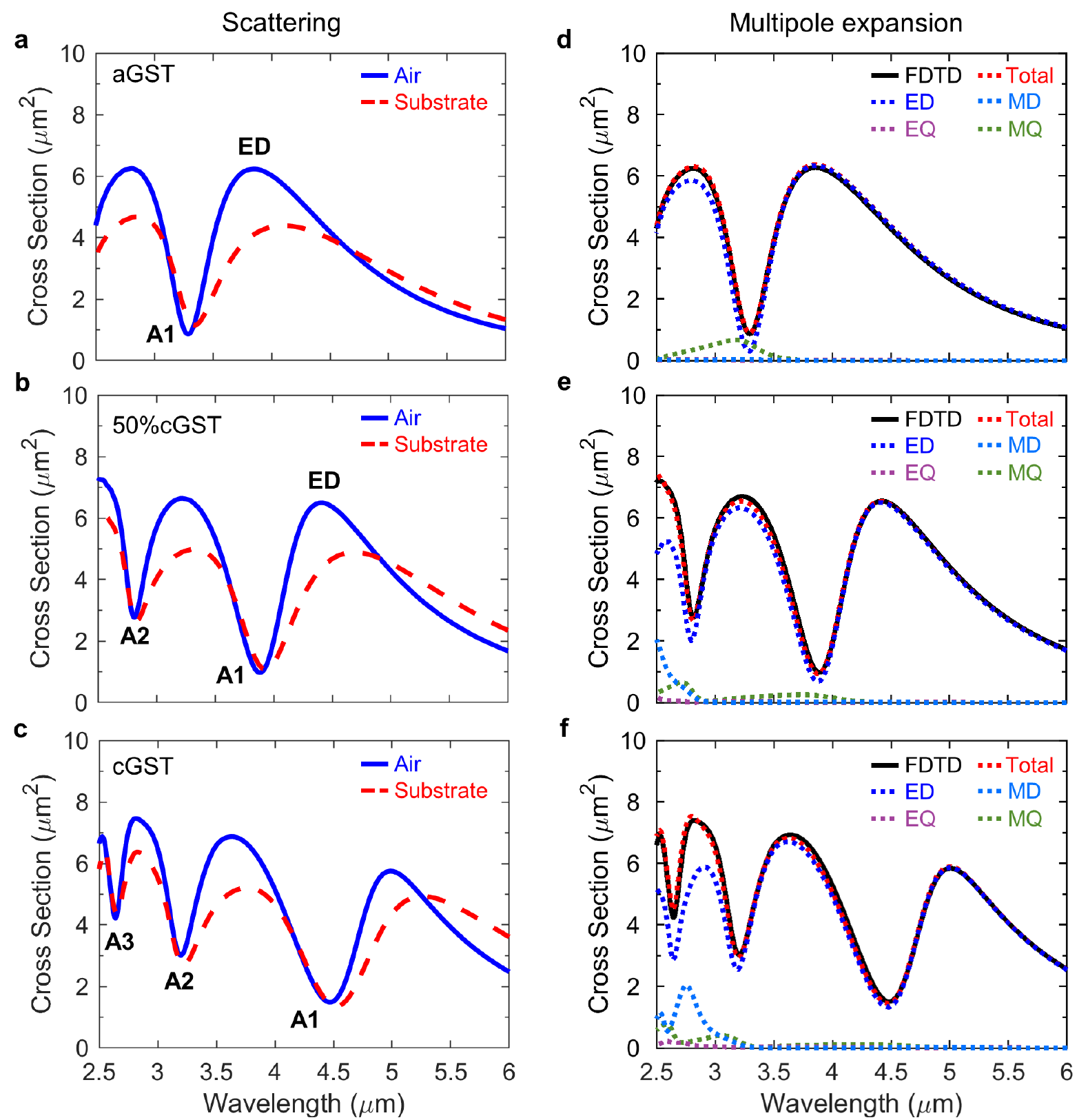}
\caption{\textbf{.} (\textbf{a}--\textbf{c}) Simulated scattering spectra of individual GST nanodisks in the vacuum and on the substrate. 
(\textbf{d}--\textbf{f}) Multipole expansion of the simulated scattering spectra. Both fundamental and higher-order multipoles can be clearly identified.}
\label{FigureS6}
\end{figure}

In Fig. S6\textbf{a}--\textbf{c}, we compare the scattering spectra with and without the substrate. Based on the Mie theory \cite{bohren2008absorption}, 
the strengths and the spectral positions of the multipolar responses are directly related to the index contrast between the dielectric material and the environment. 
Since the GST material possesses extremly high indices ($n_\text{aGST} > 4$, $n_\text{cGST} > 6$), the existence of the substrate ($n_\text{CaF$_2$} \sim 1.4$) 
thereby only has a very limited impact on the spectral positions of the scattering spectra. The associated multipole expansion in the vacuum (Fig. S6\textbf{d}--\textbf{f}) 
thus can be applied to indentify the ED and the anapole modes in the experiment. 
It is also worth noting that our multipole expansion approach (see Methods) allows us to unambiguously identify 
multipolar contributions up to arbitrarily high order.

\newrefcontext{my}
\AtNextBibliography{\small}
\printbibliography[prefixnumbers={S}]